\documentclass{article}
\usepackage{ismir,amsmath,cite,url}
\usepackage{graphicx}
\usepackage{color}
\usepackage[usestackEOL]{stackengine}
\usepackage{subcaption}
\usepackage{enumitem}

\usepackage{soul}

\newcommand{\fevd}{\texttt{FE-VD}}
\newcommand{\cnnvd}{\texttt{CNN-VD}}
\newcommand{\rnnvd}{\texttt{RNN-VD}}

\newcommand{\fevds}{\texttt{FE-VD }}
\newcommand{\cnnvds}{\texttt{CNN-VD }}
\newcommand{\rnnvds}{\texttt{RNN-VD }}

\newcommand{\squeeze}{\vspace{-0.1cm}}
\newcommand{\squeezesix}{\vspace{-0.2cm}}

\title{Revisiting Singing Voice Detection: \\A quantitative review and the future outlook}

\multauthor
{Kyungyun Lee$^1$ \hspace{0.3cm} Keunwoo Choi$^2$ \hspace{0.3cm} Juhan Nam$^3$}
{$^1$ School of Computing, KAIST \\ $^2$ Spotify Inc., USA \\ $^3$ Graduate School of Culture Technology, KAIST\\
{\tt\small \ kyungyun.lee@kaist.ac.kr, keunwooc@spotify.com, juhannam@kaist.ac.kr}
}
\def\authorname{Kyungyun Lee, Keunwoo Choi, Juhan Nam}

\begin{document}
\maketitle
\begin{abstract}
Since the vocal component plays a crucial role in popular music, singing voice detection has been an active research topic in music information retrieval. Although several proposed algorithms have shown high performances, we argue that there still is a room to improve to build a more robust singing voice detection system. 
In order to identify the area of improvement, we first perform an error analysis on three recent singing voice detection systems. Based on the analysis, we design novel methods to test the systems on multiple sets of internally curated and generated data to further examine the pitfalls, which are not clearly revealed with the current datasets. 
From the experiment results, we also propose several directions towards building a more robust singing voice detector.

\end{abstract}

\section{Introduction}\label{sec:introduction}
Singing voice detection (or VD, vocal detection) is a music information retrieval (MIR) task to identify vocal segments in a song. 
The length of each segment is typically at a frame level, for example, 100~ms. Since singing voice is one of the key components in popular music, VD can be applied to music discovery and recommendation as well as various MIR tasks such as melody extraction \cite{hsu2009singing}, audio-lyrics alignment \cite{wang2004lyrically}, 
and artist recognition \cite{berenzweig2002using}. 

%


Existing VD methods can be categorized into three different classes. \textit{First}, the early approaches focused on the acoustic similarity between singing voice and speech, utilizing cepstral coefficients \cite{berenzweig2001locating} and linear predictive coding \cite{kim2002singer}. The \textit{second} class would be the majority of existing methods, where the systems take advantages of machine learning classifiers such as support vector machines or hidden Markov models, combined with large sets of audio descriptors (e.g., spectral flatness) as well as dedicated new features such as fluctograms \cite{lehner2014reduction}. \textit{Lastly}, there is a recent trend towards feature learning using deep neural networks, with which the VD systems learn optimized features for the task using a convolutional neural network (CNN) \cite{schluter2015exploring} and a recurrent neural network (RNN) \cite{leglaive2015singing}. They have achieved state-of-the-art performances on commonly used datasets with over 90\% of the true positive rate (recall) and accuracy.

We hypothesize that there are common problems in existing VD methods in spite of such well-performing metrics that have been reported. Our scope primarily includes methods in the second and third classes since they significantly outperform those in the first class. Our hypothesis was inspired by inspecting the assumptions in the existing algorithms. The most common one, for example, has been made on the spectro-temporal characteristics of singing voices; that they include frequency modulation (or vibrato) \cite{markaki2008singing, regnier2009singing}, which leads to our analysis on whether there are any problems by pursuing to be a vibrato detector. We can also raise similar questions on the behavior of the systems in the third class, the deep learning-based systems, by examining on their assumptions and results. 
Based on the analysis, we invent a set of empirical analysis methods and use them to reveal the exact types of problems in the current VD systems.


\begin{table*}
\centering
\setlength\tabcolsep{1.0pt} 
 \begin{tabular}{c||c|c|c|c}
  \hline
   & Size & Annotations & Past VD papers & Notes \\
  \hline
  \hline
  Jamendo Corpus  & 93 tracks (443 mins) & Vocal activation & {\Centerstack{\cite{leglaive2015singing}, \cite{regnier2009singing},\cite{lehner2013towards},\\ \cite{lehner2015low}, \cite{schluter2015exploring}, \cite{schluter2016learning}}} & {\Centerstack{ Train/valid/test split from \cite{ramona2008vocal}}}\\
  \hline
  RWC Popular Music  & 100 tracks (407 mins) & {\Centerstack {Vocal activation,\\ instrument annotation}} &{\Centerstack{\cite{schluter2016learning},\cite{schluter2015exploring},\cite{lehner2014reduction} \\ \cite{lehner2015low}}}& VD annotation by \cite{mauch2011timbre} \\
  \hline
  MIR-1K & 100 short clips (113 mins) & {\Centerstack {Vocal activation, \\ pitch contours}} &\cite{hsu2012tandem} & {\Centerstack {Regular speech files \\ provided}} \\
  \hline
  MedleyDB & {\Centerstack{122 tracks (437 mins)}} &{\Centerstack{ Melody annotation, \\ pitch annotation}} &\cite{schluter2016learning}& Multitrack \\
  \hline
 \end{tabular}%
 \caption{A summary of public datasets relevant to singing voice detection}
 \label{tab:dataset}
\end{table*}

Our contributions are as follows :  \squeeze
\begin{itemize}[leftmargin=*]
\item A quantitative analysis to clarify and classify common errors of three recent VD systems (\secref{sec:experiment1})
\squeeze
\item An analysis using curated and generated audio contents that exploit the discovered weakness of the systems (\secref{sec:experiment2})
\squeeze
\item Suggestions on future research directions (\secref{sec:directions})
\end{itemize}
\squeeze
In addition, we review previous VD systems in \secref{sec:models} and summarize the paper in \secref{sec:conclusion}. 

\section{Background}
\subsection{Problem definition}\label{subsec:probdef}
Singing voice detection is usually defined as a binary classification task about whether a short audio segment input includes singing voice. However, the details have been rather empirically decided. By `short', the segment length for prediction is often 100~ms or 200~ms. `Audio' can be provided as stereo, although they are frequently downmixed to mono. More importantly, `singing voice' is not clearly defined, for example, leaving the question that background vocals should be regarded as singing voice or not. In previous works, this problem has been neglected since the majority of songs in datasets do not include background vocals that are independent of the main vocals. These will be further discussed in \secref{sec:directions}.

\subsection{Public Datasets}\label{subsec:datasets}

In \tabref{tab:dataset}, four public datasets for evaluating VD systems are summarized. Three of them are well described by Lehner et al.\cite{lehner2013towards}: Jamendo Corpus\cite{ramona2008vocal}, RWC Popular Music Database\cite{goto2002rwc} and MIR-1K Corpus\cite{hsu2010improvement}. In addition, we add MedleyDB\cite{bittner2014medleydb}, which is a multitrack dataset, composed of raw mono recordings for each instrument as well as processed stereo mix tracks. Although it does not provide annotations for vocal/non-vocal segments, it is possible to utilize the annotations for the instrument activation, which considers vocals as one of the instruments. There can be more benefit by using the multitrack dataset for VD research, which will be discussed in \secref{sec:directions}.

\squeeze

\subsection{Audio Representation} \label{ssec:audiorepre}
In this section, we present the properties as well as the underlying assumptions of various audio representations in the context of VD. Previous works have used a combination of numerous audio features, seeking easier ways for the algorithm to detect the singing voice. They range from representations such as short-time Fourier transform (STFT) to high-level features such as onsets and pitch estimations. 


\squeeze

\begin{itemize}[leftmargin=*]
\item \textbf{STFT} provides a 2-dimensional representation of audio, decomposing the frequency components. 
STFT is probably the most basic (or `raw') representation in VD, based on which some other representations are either designed and computed, or learned using deep learning methods. 
\squeeze

\item \textbf{Mel-spectrogram} is a mel-scaled frequency representation and usually more compressive than STFTs and originally inspired by the human perception of speech. 
Being closely related to speech provides a good motivation to be used in VD, therefore mel-spectrogram has been actively used as an input representation of CNNs \cite{schluter2015exploring} and RNNs \cite{leglaive2015singing}. When deep learning methods are used, mel-spectrogram is often preferred due to its efficiency compared to STFT. 
\squeeze


\item \textbf{Spectral Features} such as spectral centroid and spectral roll-off are statistics of a spectral distribution of a single frame of time-frequency representations (e.g., STFT). A particular and most noteworthy example is \textbf{Mel-Frequency Cepstral Coefficients} (MFCCs). MFCCs have originally been designed for automatic speech recognition and take advantages of mel-scale and fourier analysis for providing approximately pitch-invariant timbre-related information. They are often (assumed to be) relevant to MIR tasks including VD \cite{rocamora2007comparing,lehner2013towards}. Spectral features, in general, are not robust to additive noise, which means that they would be heavily affected by the instrumental part of the music when used for VD.

\end{itemize}


\squeeze
\squeeze

\section{Models}\label{sec:models}
\begin{figure}[t]
\centering
 \includegraphics[width=\columnwidth]{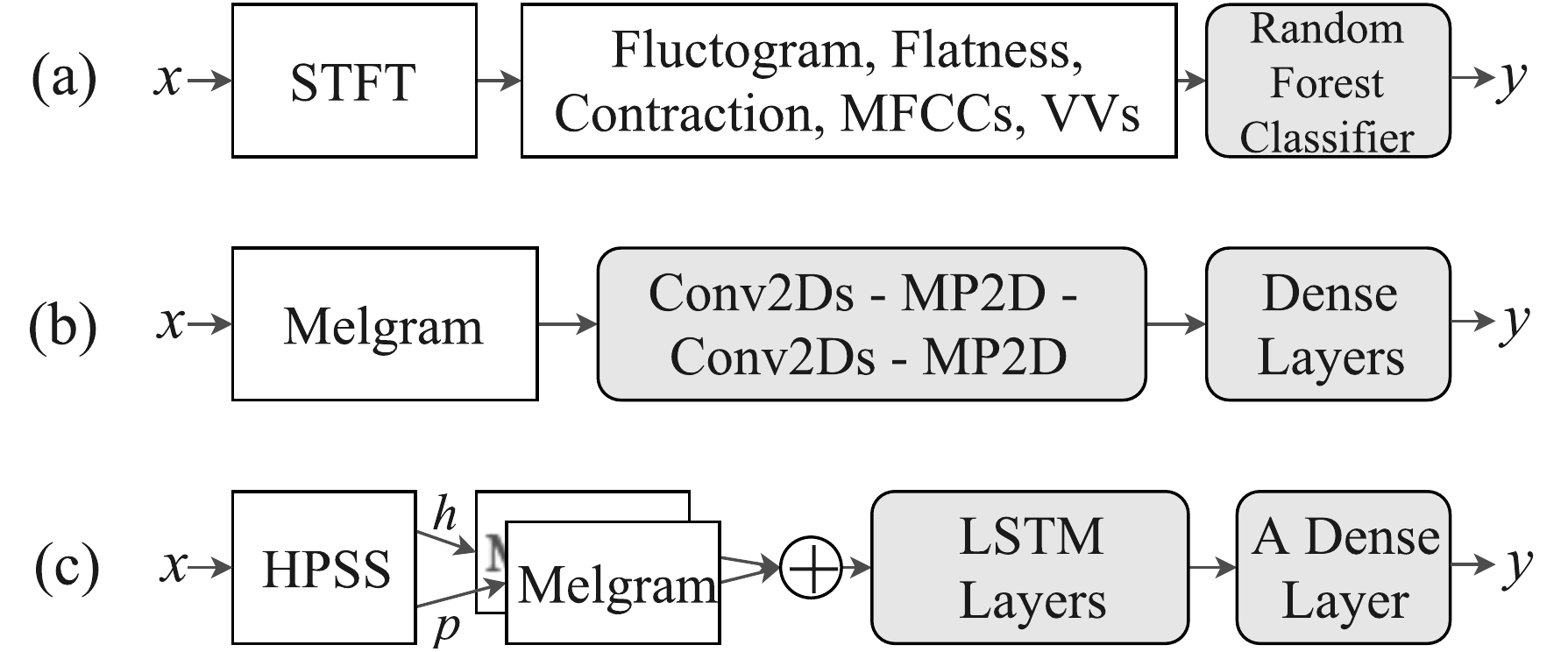}
  \caption{Block diagrams for three VD systems -- (a) \fevd \cite{lehner2014reduction}, (b) \cnnvd \cite{schluter2015exploring}, and (c) \rnnvd \cite{leglaive2015singing}. \textit{x} and \textit{y} for input audio signal and output prediction (probability of singing voice). Rounded, gray blocks are trainable classifiers or layers. The details of the features in (a) are explained in \cite{lehner2014reduction}. In (c), `+' indicates frequency-axis concatenation and `h' and `p' are the separated harmonic/percussive components. 
  } 
  \label{fig:diagrams}
\end{figure}

In this section, we introduce three recent and distinctive VD systems that have improved the state-of-the-art performances 
along with the details of our re-implementation of them.\footnote{http://github.com/kyungyunlee/ismir2018-revisiting-svd} 
They are briefly illustrated in Figure \ref{fig:diagrams}, where $x$ and $y$ indicate the input audio signal and prediction respectively.


\squeeze

\subsection{Lehner et al.\cite{lehner2014reduction} (\fevd)}\label{ssec:fevd}
This feature engineering (FE) method, \fevds is based on fluctogram, spectral flatness, vocal variance and other hand-engineered audio features. We select this model for its rich and task-specific feature extraction process to compare with the other models. Although the features are ultimately computed frame-wise, `context' from the adjacent frames are taken into account, supposedly enabling the system to use dynamic aspect of the features. The features are aimed to reduce the false positive rate caused by the confusion between singing voice and pitch-varying instruments such as woodwinds and strings. Random forest classifier was adopted as a classifier,  
achieving an accuracy of 88.2\% on the Jamendo dataset.
While their methods have shown reduction in the false positive rates on strings, Lehner et al. mentions woodwinds such as pan flutes and saxophones still show high error rate.

Same as in \cite{lehner2014reduction}, we extract 6 different audio features (fluctograms, spectral flatness, spectral contraction, vocal variances, MFCCs and delta MFCCs), resulting in 116 features per frame. We use input size of 1.1~seconds as the input to the random forest classifier, where we performed grid search to find optimal parameters. As a post processing step, we apply the median filter of 800~ms on the predictions. 

\subsection{Schl{\"u}ter et al.\cite{schluter2015exploring} (\cnnvd)} \label{ssec:cnnvd}
Recently, VD systems using deep learning models have shown the state-of-the-art result \cite{schluter2015exploring,schluter2016learning,leglaive2015singing}. These systems often use basic audio representations such as STFT as an input to the system such as CNN and RNN, expecting the relevant features are learned by the model. We first introduce a CNN-based system \cite{schluter2015exploring}.

Schl{\"u}ter et al. suggested a deep CNN architecture with 3-by-3 2D~convolution layers. We name the CNN model \cnnvd. As a result, the system extracts trained, relevant \textit{local} time-frequency patterns from its input, a mel-spectrogram. During training, they apply data augmentation such as pitch shifting and time stretching on the audio representation. They reported that it reduces the error rate from 9.4\% to 7.7\% on the Jamendo dataset.

Our CNN architecture is identical to the original one in using an input size of 115 frames (1.6~sec) and using 4 3$\times$3 2D convolutional layers. However, we did not perform data augmentation for a fair comparison with other models. Here, we also apply the median filter of 800~ms. 


\squeeze

\subsection{Leglaive et al.\cite{leglaive2015singing} (\rnnvd)} \label{ssec:rnnvd}
As another deep learning-based system, Leglaive et al. \cite{leglaive2015singing} proposed a recurrent neural network with bi-directional long short-term memory units (Bi-LSTMs) \cite{hochreiter1997long}, with an assumption that temporal information of music can provide valuable information for detecting vocal segments. We name this system \rnnvd. For the classifier input, the system performs double-stage harmonic-percussion source separation (HPSS) \cite{ono2008separation} on the audio signal to extract signals relevant to the singing voice. 
For each frame, mel-spectrograms of the obtained harmonic and percussive components are concatenated as an input for the classifier. Several recurrent layers followed by
a shared densely-connected layer (also known as time distributed dense layer) 
yield the output predictions for each input frame.
This model achieves the state-of-the-art result without data augmentation, showing accuracy of 91.5\% on the Jamendo dataset. From this result, although the contributions from additional preprocessing vs. recurrent layers may be combined, we can assume that past and future temporal context help to identify vocal segments.

For our RNN architecture, we use the best performing model from the original article\cite{leglaive2015singing}, one with three hidden layers of size 30, 20 and 40. The input to the model is 218 frames~(3.5 seconds).

\squeeze

\section{Experiment I: Error categorization}\label{sec:experiment1}

The purpose of this experiment is to identify common errors in the VD systems through our implementation of models from \secref{sec:models}. The results and observations lead to the motivation of experiments in \secref{sec:experiment2}. Librosa\cite{mcfee2017librosa} is used in audio processing and feature extraction stages. 

\squeeze

\subsection{Data and Methods}
Three systems (\fevd, \cnnvd, \rnnvd) are trained on the Jamendo dataset with a suggested split of 61, 16 and 16 for train, validation and test sets \cite{ramona2008vocal}, respectively.
They are primarily tested on the Jamendo test set. For qualitative analysis, we also utilize MedleyDB. 
Note that MedleyDB does not provide vocal segment annotations, so we use the provided annotation for instrument activation to create ground truth labels for vocal containing songs. 



\begin{table}
\begin{center}
 \begin{tabular}{c||c|c|c}
   & \fevd & \cnnvd & \rnnvd \\
  \hline
  \hline
  Acc.(\%)  & 87.9 & 86.8 & 87.5 \\
  \hline
  Recall(\%) & 91.7 & 89.1 & 87.2 \\
  Precision(\%) & 83.8 & 83.7 & 86.1 \\
  F-measure(\%) & 87.6 & 86.3 & 86.6 \\
  \hline 
  FPR(\%) & 15.3 & 15.1 & 12.2 \\
  FNR(\%)& 8.3 & 10.9 & 12.8 \\
 \end{tabular}
\end{center}
\squeeze
\squeeze

\caption{Results of our implementations on the Jamendo test set. FPR and FNR refer to false positive rate and false negative rate, respectively.}
\label{tab:jamendo}
\end{table}

\squeeze

\subsection{Results}
The test results of our implementation are shown in \tabref{tab:jamendo}. We did not focus on fine-tuning individual models because three systems altogether are used as a tool to get a generalized view of the recent VD systems, thus showing slightly lower performances compared to the results in original papers. 
Overall, \fevd, \cnnvds and 
\rnnvds show a negligible difference on the test scores.
We observe trends that are similar to the original papers in terms of performance and the precision/recall ratio. 

Upon listening to the misclassified segments, we categorize the source of errors into three classes -- pitch-fluctuating instruments, low signal-to-noise ratio of the singing voice, and non-melodic sounds. 
%

\begin{table}[t]
\centering
\label{songs}
\setlength\tabcolsep{2.0pt} 
\begin{tabular}{l|c|c|c|c}
\small Song Title & \small Confusing inst & \small \fevd & \small \cnnvd & \small \rnnvd \\ \hline
 \hline
\small LIrlandaise & \small Woodwind, Synth & 46.6 & 29.5 & 22.0 \\ 
\small Castaway & \small Elec. Guitar & 62.5 & 56.5 & 24.2 \\ \hline \hline
\small Say me Good Bye  & \small N/A  & 2.8 & 3.0 & 2.5  \\ 
\small Inside &  \small N/A & 5.9 & 6.7 &  5.0 \\ 
\end{tabular}
\caption{False positive rate (\%) of each system for 4 songs from the Jamendo test set. The top 2 songs are the ones ranked within the top 5 lowest accuracy and the bottom 2 songs are the ones ranked within the top 5 highest accuracies at song level across all three systems.}
 \label{tab:songs}
\end{table}

\subsubsection{Pitch-fluctuating instruments} \label{sssec:error_pitch_fluc}
Classes of instruments such as strings, woodwinds and brass exhibit similar characteristics as the singing voice, which we refer to as being `voice-like'\cite{schubert2016voicelikeness}. By `voice-like', we consider three aspects of the signal, namely, pitch range, harmonic structure, and temporal dynamics (vibrato). Especially, we find temporal dynamics as important attributes that are recognized by the VD systems to identify vocal segments.

\begin{figure}[t]
\centering
 \includegraphics[width=\columnwidth]{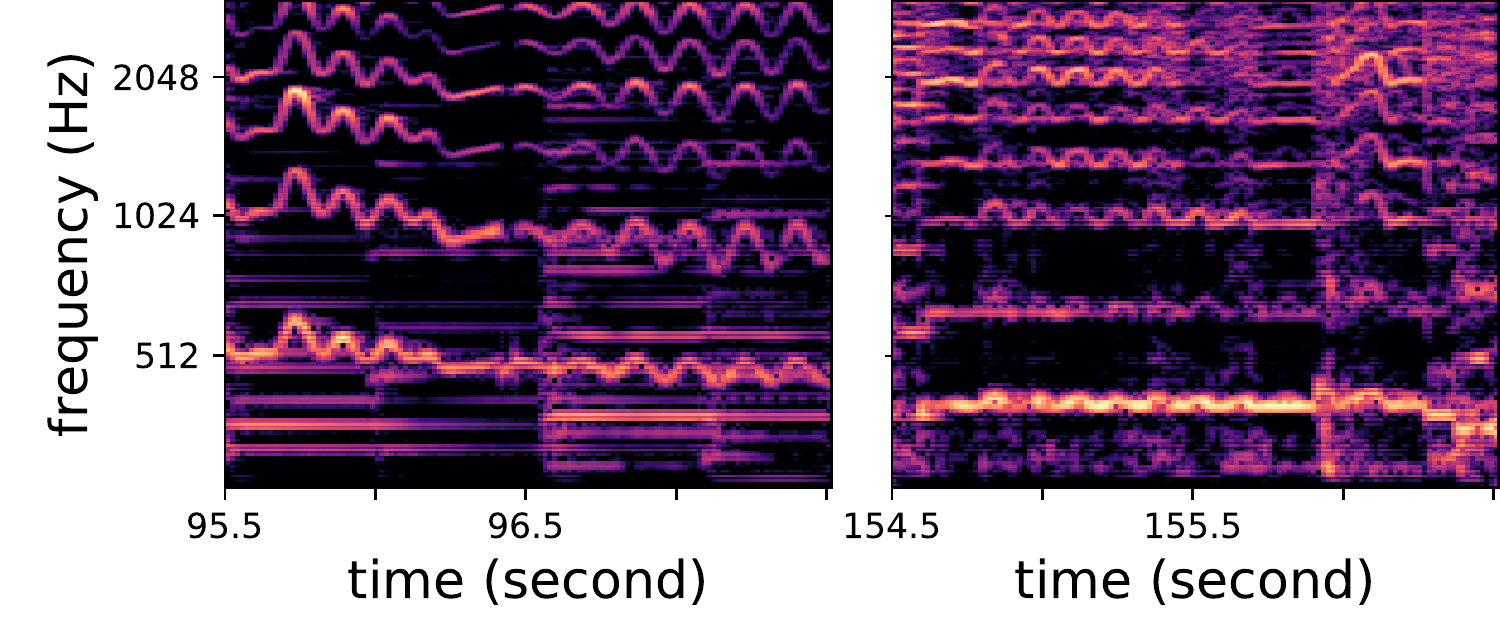}
  \caption{Excerpts of Mel-spectrograms from MedleyDB: `Handel\_TornamiAVagheggiar' with female vocalist (left) and `PurlingHiss\_Lolita' with electric guitar (right) (see \secref{sssec:error_pitch_fluc}.)}
  \label{fig:CNNmel}
\end{figure}

Frequency modulation, also known as vibrato, resembles the modulation created from the vowel component of singing voice. This is illustrated in \figref{fig:CNNmel}, where mel-spectrograms of both a female vocalist and an electric guitar show curved lines. 
We observe that this similarity causes further confusion in the system. 

In \tabref{tab:songs}, we list two songs found among the top 5 least/most accurately predicted songs in the test set of all three systems. The woodwind in `05 - Llrlandaise' causes high false positives, which may be due to the presence of vibrato and the similarity in pitch range to that of soprano singers (above 220~Hz). \fevds and \cnnvds show poor performance on woodwinds, probably because the fluctogram of \fevds and small 2D convolution kernels of \cnnvds are specifically designed to detect vibrato as one of the features for identifying singing voice. In the same song, all three systems show confusion with the synthesizer. Synthesizers mimicking pitch-fluctuating instruments are particularly challenging as it is difficult to characterize them as specific instrument types.        

In addition, electric guitars are one of the most frequently found sources of false positives, as can be seen from `03 - castaway', mostly caused by the recognizable vibrato patterns. We find the confusion gets worse when the guitar is played with effects, like wah-wah pedals, which imitates the vowel sound of the human. Lastly, we note that some of the other problematic instruments in our test sets include saxophones, trombones and cellos, which are well-known `voice-like' instruments. 


This observation, regarding the system pitfalls on vibrato patterns, is further investigated in \secref{ssec:vibrato}.

\squeeze

\subsubsection{Signal-to-noise ratio and the performance}\label{sssec:SNRPerf}
Lastly, we note that all the three systems are affected by the signal-to-noise ratio (SNR), or the relative gain of vocal component, as one can easily expect. All of the three systems exhibit high false negative rate when the vocal signal is relatively at a low level. 

In systems such as Lehner et al., where audio features such as MFCCs or spectral flatness are used, the performance varies by SNR because the features are statistics of the whole bandwidth which includes not only the target signal (vocal) but also additive noise (instrumental). VD systems with deep neural networks are not free from this since the low-level operation in the layers of deep neural networks are a simple pattern matching by computing correlation.

This is a common phenomenon in other tasks as well, e.g., speech recognition, and we continue the discussion to a follow-up experiment in \secref{ssec:audiogain} and finally a suggestion on the problem definition and dataset composition in \secref{sec:directions}.

\subsubsection {Non-melodic Sources} \label{sssec:error_voices}
Although the interest of most VD systems appears to lie mainly in the melodic component of the song, we expected the system to learn percussive nature of the singing voice as well, which is exhibited by consonants from the singers. Therefore, our hypothesis is whether the system is confused by the consonants of singing voice and percussive instruments, resulting in either \textit{i)} missing consonant parts (false negative) or \textit{ii)} mis-classifying percussive instruments (false positive). 

From our test results, we encounter false positive segments containing snare drums and hi-hats, but the exact cause of this misclassification is unclear. We further tested the system with drum set solos for potential false positives and with a collection of consonant sounds such as plosives and fricatives from the human voice for potential false negatives, but we did not observe a clear pattern in misclassification. Although we do not conduct further experiment on this, it suggests a deeper analysis, which may also lead to a clear understanding of preprocessing strategies including HPSS. 



\squeeze

\section{Experiment II: Stress testing}\label{sec:experiment2}



\subsection{Testing with artificial vibrato}\label{ssec:vibrato}
Based on the confusion between `voice-like' instruments and singing voice, we hypothesize that the current VD systems use vibrato patterns as one of the main tools for vocal segment detection. We explore the degree of confusion for each VD system by testing them on synthetic vibratos with varying rate, extent and formant frequencies. 




\subsubsection{Data Preparation}
We create a set of synthetic vibratos with low pass-filtered sawtooth waveforms with $f_0$=220~Hz. We vary the modulation rate and frequency deviation ($f_\Delta$) to investigate their effects. Furthermore, we apply 5~bi-quad filters at the corresponding formant frequencies (3 for each) to synthesize so that they would sound like the basic vowel sounds, `a', `e', `i', `o', `u' \cite{smith2007introduction}. The modulation rate ranges in \{0.5, 1, 2, 4, 6, 8, 10~Hz\} and the frequency deviation ranges in \{0.01, 0.1, 0.3, 0.6, 1, 2, 4, 8 semitones\} with respect to its $f_0$). As a result, the set consists of $7$ (rates) $\times 8$ ($f_\Delta$'s) $\times 6$ (5~formants + 1~unfiltered) $=336$ variations. 
\subsubsection{Results}

\begin{figure}[t]
\centering
 \includegraphics[width=\columnwidth]{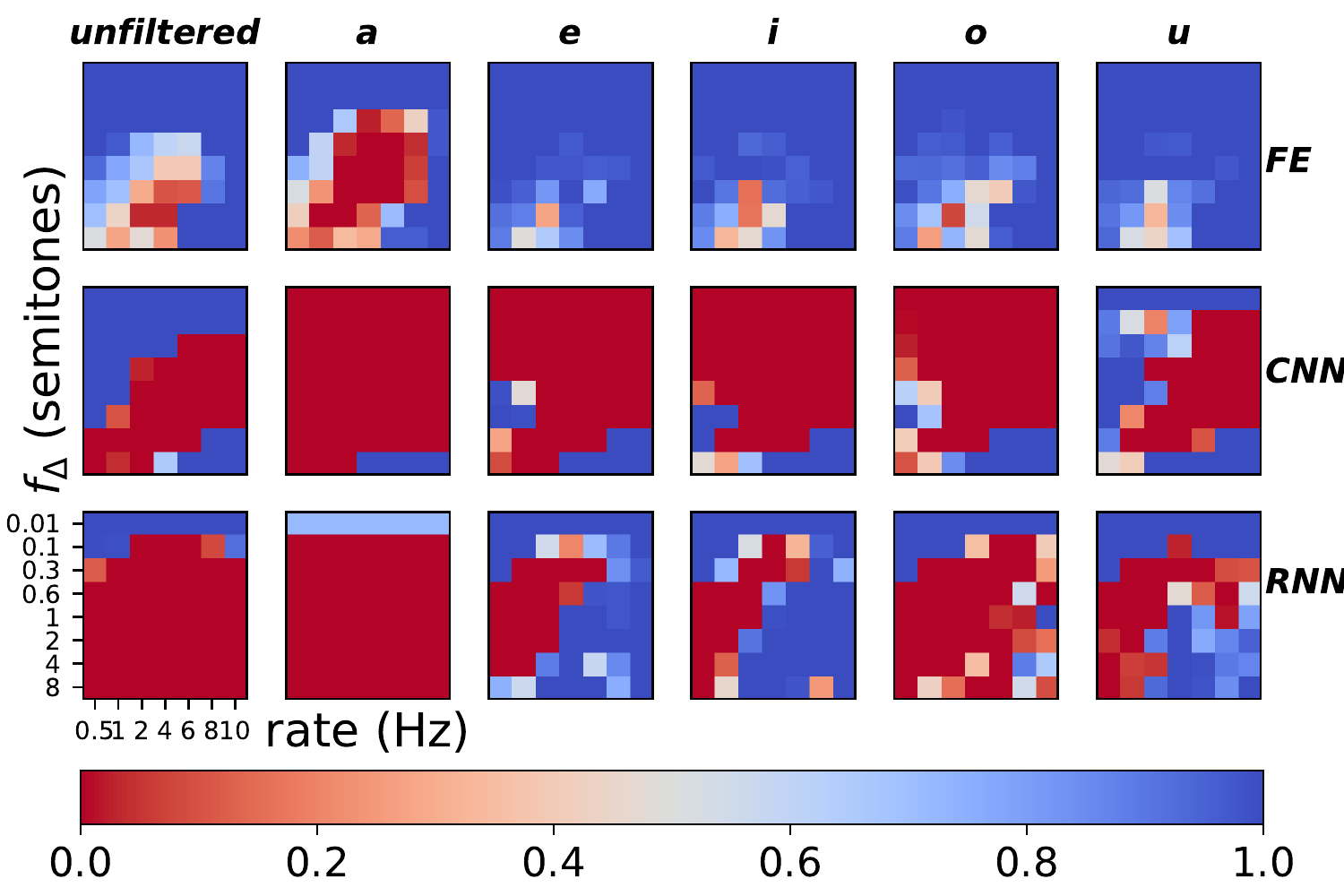}
  \caption{Heat-maps of the accuracies of the vibrato experiment result. Each row corresponds to VD systems (\fevd, \cnnvd, \rnnvd) and each column corresponds to the formant (unfiltered, 'a', 'e', 'i', 'o', 'u'). Within each heat map, x- and y-axes correspond to the vibrato rate and frequency deviation as annotated on the lower-left subplot (see \secref{ssec:vibrato})}
  \label{fig:vibrato}
\end{figure}

\figref{fig:vibrato} shows the result of the prediction by the three VD systems on the synthetic vibratos. The accuracy of 1.0 indicates that the system does not confuse the artificial vibratos with singing voice. Here, we observe the performance difference of each model, which were not visible from looking at the scores in \tabref{tab:jamendo}. In general, confusion areas tend to be concentrated on the bottom left to the center area of the graph. The extent and rate of the artificial tones that are highly misclassified seem to be around the range of vibratos of singers, which is said to be around 0.6 to 2 semitone with rate around 5.5 to 8~Hz\cite{timmers2000vibrato}. We also observe a within-system difference, i.e., the presence and the type of formants affect the models. For instance, vibratos mimicking the vowel `a' cause higher misclassification in all three models.  


\fevds performs much better than the latter two systems. Note that \fevds is a feature engineering model, where unique features, such as fluctogram and vocal variance, are mostly adapted from the ones used in speech recognition task. As these features were intentionally designed to reduce false positives from pitch-varying instruments, it appears to significantly reduce error rate on vibratos with rate and extent that are beyond the range of human singers.  

\cnnvds confuses slightly wider range of vibratos. This is expected to some extent since the model prominently uses 3$\times$3 filters on mel-spectrogram to detect local features, which can be regarded as a \textit{local} pattern detector. In other words, the locality of CNN results in a system that is easily confused by frequency modulation regardless of the non-singing voice aspects of the signal. This implies that the model may benefit from looking at a varying range of time and frequency to learn vocal-specific characteristics, such as timbre \cite{pons2017timbre}. 

Lastly, \rnnvds performs better than the \cnnvd, though worse than \fevd. On detecting vocal and non-vocal segments, it seems natural, even for humans, that past and future temporal context help. Also, we presume that the preprocessing of double stage HPSS contributes to the robustness of the system against vibrato. Again, this observation leaves a question of separating the contributions from preprocessing and model structure.  



\begin{figure*}[t]
  \begin{subfigure}[b]{0.33\linewidth}
    \includegraphics[width=\columnwidth]{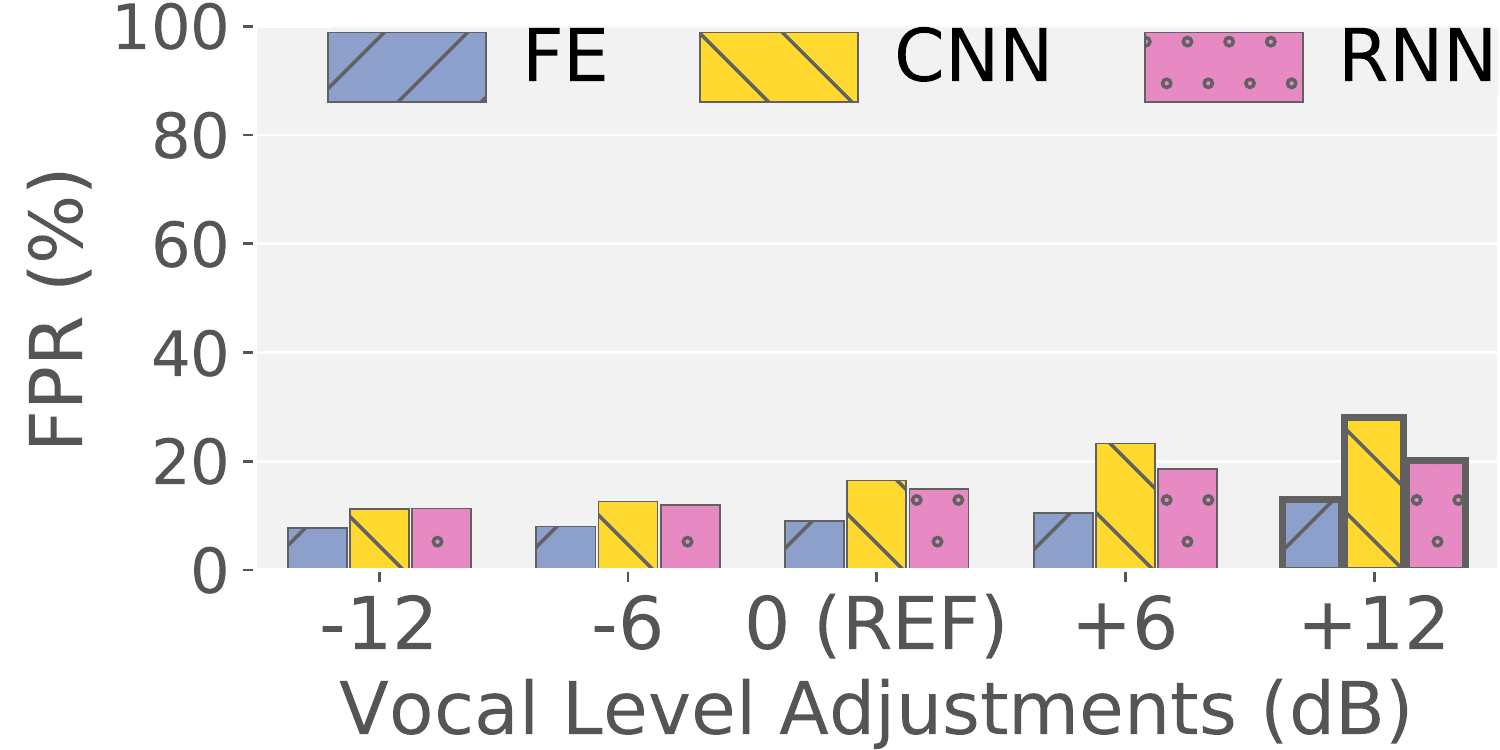}
  \end{subfigure}
  \hfill 
  \begin{subfigure}[b]{0.33\linewidth}
    \includegraphics[width=\columnwidth]{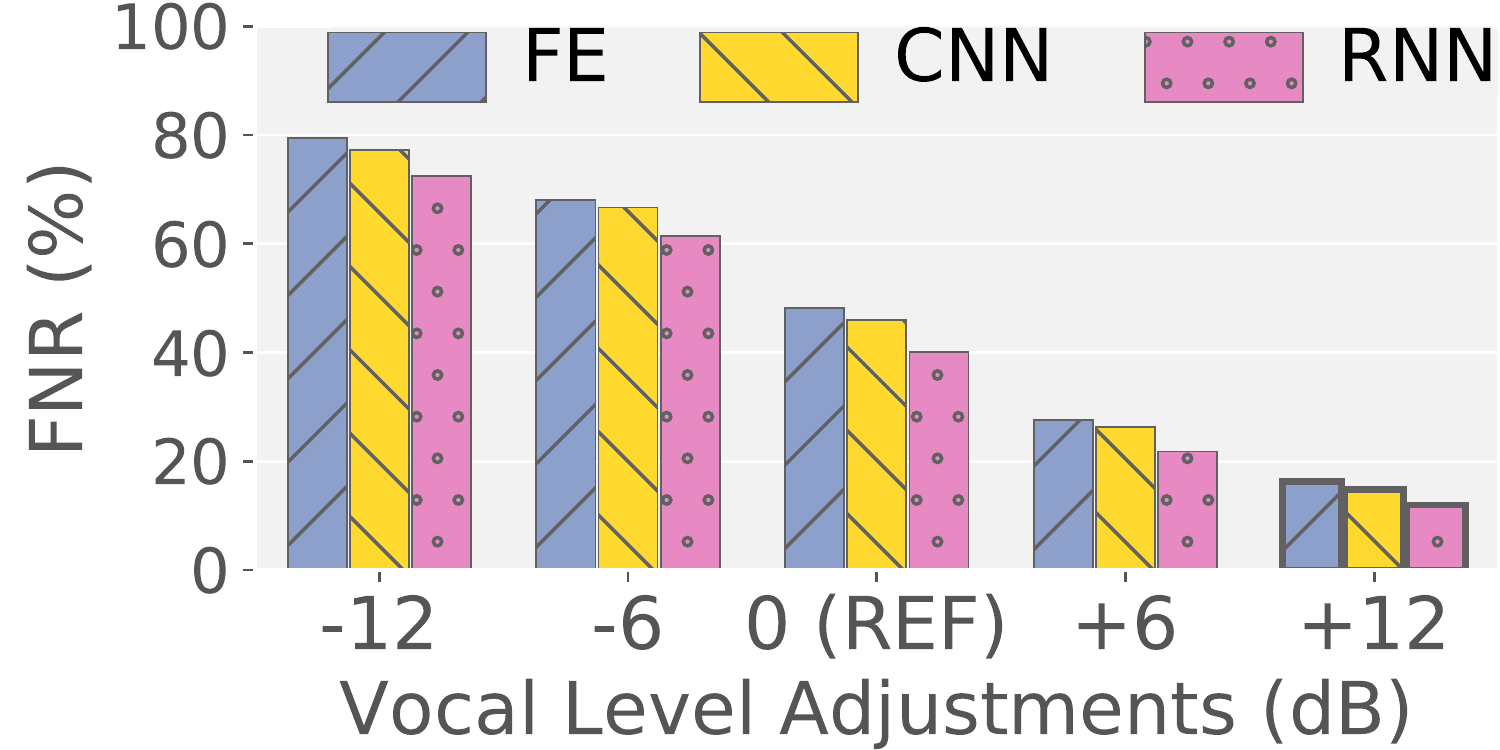}
  \end{subfigure}
    \hfill 
  \begin{subfigure}[b]{0.33\linewidth}
    \includegraphics[width=\columnwidth]{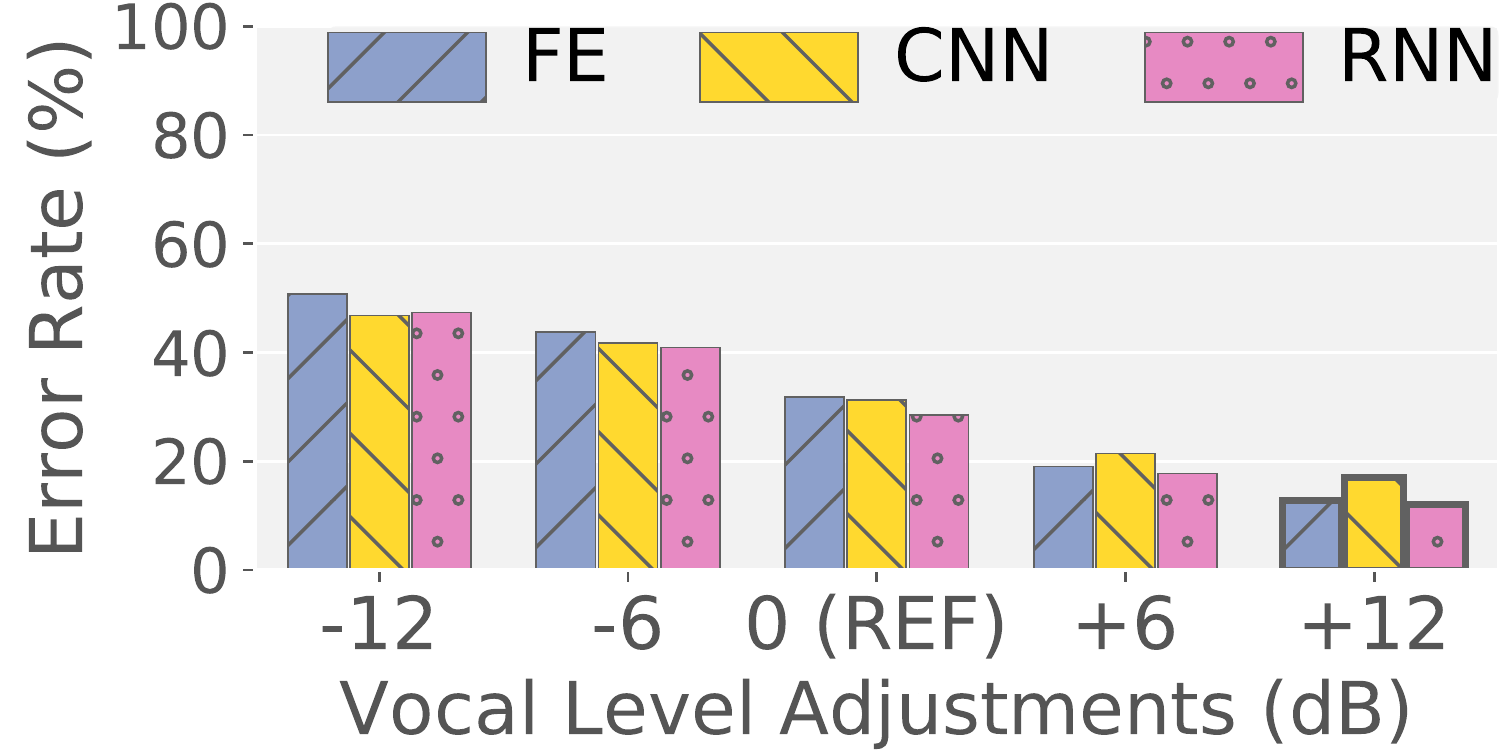}
  \end{subfigure}
  \caption{False positive rates, false negative rates, and overall error rates for the three systems in the stress testing with controlling SNR (see \secref{ssec:audiogain}).}
  \label{fig:loudness}
  \squeezesix
\end{figure*}

\subsection{Testing with SNR}\label{ssec:audiogain}
\squeeze
In this experiment, VD systems are tested with vocal gain adjusted tracks to further explore the behavior of the systems on various scenarios, which can reflect the real-world audio settings of live recordings and radios, for example.
\squeeze

\subsubsection{Data preparation}
We create a modified test set using 61 vocal-containing tracks provided by MedleyDB. We use the first 30~seconds of the songs to build a pair of (vocal, instrumental) tracks. Vocal tracks are modified with SNR of \{+12 dB, -12 dB, +6 dB, -6 dB, 0 dB\}.

\squeeze

\subsubsection{Results}
The results of the energy level robustness test are presented in \figref{fig:loudness} with false positive rate, false negative rate, and overall error rate. We see a consistent trend across the performance of all three VD systems, which is once again an expected pattern as aforementioned in \secref{sssec:SNRPerf} -- that increasing SNR help to reduce false negatives. Overall error rate also exhibits a noticeable decrease in common with higher SNRs. In practice, one could take advantage of data augmentation with changing SNR to build a more robust system. More importantly, it can be part of the evaluation procedure for VD, as we discuss in \secref{sec:directions}.

While the VD systems behave similarly on all test cases, we note that \fevd, owing to its additional features, shows lowest variance and lowest value for the false positive rate. 
Also, our assumption that the double-stage HPSS, which filters out vocal-related signals, would make \rnnvds more robust against SNR is observed to be not necessarily true as we clearly see performance differences across the varying SNR test cases.

\section{Directions to improve} \label{sec:directions}
\subsection {Defining the problem and the datasets}

\subsubsection{Defining singing voice}
By using the annotations in datasets such as Jamendo, many VD systems implicitly assume that the target `singing voice' is defined as vocal components that correspond to the \textit{main melody}. Other voice-related components such as backing vocal, narration, humming, and breathing are not clearly defined to be singing voice or not.

In some applications, however, they can be of interest. For example, a system may want to find purely instrumental tracks, avoiding tracks with backing vocal. In this case, the method should consider backing vocal as singing voice. However, for a Karaoke application, only the singing voice of the main melody would matter.

Therefore, an improvement can be made on defining the VD problem and creating datasets. For the annotation, a hierarchy among the voice-related components can be useful for both structured training and evaluation of a system \cite{mcfee2017structured, DBLP:conf/cvpr/RedmonF17}. For the audio input, we see a great benefit of multitracks, where main vocal melody, backing vocal, and other components are provided separately.


\subsubsection{Varying-SNR scenarios}
For a long while, varying SNR had been one of the common ways to evaluate speech recognition or enhancement using dataset such as Aurora \cite{hirsch2000aurora}. As observed in \secref{sssec:SNRPerf}, it can be used as a `test-set augmentation' to measure the performance of a system more precisely. Also, it can be an additional data augmentation method along with the ones in \cite{schluter2015exploring} to build a VD system more robust to various audio settings, such as audios from user generated videos. These can both be easily achieved with a multitrack dataset in practice.

\subsubsection{Measuring dataset noise}
Human annotators are neither perfect or identical, thus causing annotation noise and disagreement. Since VD is a binary classification problem, we may remain optimistic by assuming that the annotation noise is a matter of temporal precision, which is arbitrary and not agreed among many datasets so far. For example, in RWC Popular Music \cite{mauch2011timbre}, ``short background segments of less than 0.5-second duration were merged with the preceding region'' and the annotations have 8 decimal digits (in second), while in Jamendo, they are 3 decimal digits. The optimal precision may depend on human perception of sound which is often said around 10~ms in general \cite{moore2012introduction}. Although it would require a deeper investigation, the current temporal precision may be too high, leading to evaluate the systems with an overly precise annotation.




\squeezesix

\subsection{Learning from human perception}

The characteristic of voice was the main motivation in the very early works exploiting speech-related features \cite{kim2002singer,berenzweig2001locating}. Clearly, however, those approaches that solely relied on speech features showed limited performances. While following researches have improved the performance, as our experiments have demonstrated through this paper, the systems do not completely take advantage of the cues that human is probably using, e.g., the global formants, linguistic information, musical knowledge, etc. 







\squeezesix
\subsection{Preprocessing}
A light-weight VD system was introduced in \cite{lehner2013towards} where only MFCCs were used to achieve a precision of 0.788 on Jamendo dataset. This implies that there is a possibility to achieve better performance by optimizing the preprocessing stage. One of the unanswered questions is the effect of the preprocessing stage in \rnnvd \cite{leglaive2015singing} as well as whether similar processing could lead to better performance with other systems, e.g., CNN \cite{schluter2015exploring}.





\squeezesix
\section{Conclusions}\label{sec:conclusion}
In this paper, we suggested that there still are several areas to improve for the current singing voice detectors. In the first set of experiments, we identified the common errors through error analysis on three recent systems. Our observations that the main sources of error are pitch-fluctuating instruments and low signal-to-noise ratios of the singing voice motivated us to further perform stress tests. Testing with synthetic vibratos revealed that some systems (\fevd) are more robust to non-vocal vibratos than others (\cnnvds and \rnnvd). SNR-varying test showed that SNR manipulation greatly affects the current VD systems, thus it can potentially be used to strengthen the VD systems to become invariant to a wider range of audio settings. 
As we propose several directions for a more robust singing voice detector, we note that defining the VD problem is dependent on the goal of the system, thus using multitrack datasets can be beneficial. Our future interest is to further investigate on SNR to extend VD systems on uncontrolled audio settings and to examine different components of individual systems, including the preprocessing stage.

\section{Acknowledgements}
We thank Bernhard Lehner and Simon Leglaive for active discussion and code, Jeongsoo Park for sharing Ono's code. This work was supported by the National Research Foundation of Korea (Project 2015R1C1A1A02036962).

\bibliography{ISMIRtemplate}


\end{document}